\documentclass[submission,copyright]{eptcs}
\providecommand{\event}{Wivace 2013}
\usepackage{breakurl}
\usepackage{amsmath}
\usepackage{graphicx}
\usepackage{subfig}
\usepackage{amssymb}

\title{An ensemble approach to the study of the emergence of metabolic and proliferative disorders via\\ Flux Balance Analysis}

\author{Chiara Damiani$^a$ $^b$ \quad Riccardo Colombo$^a$ $^b$\quad Sara Molinari$^a$ $^c$\quad Dario Pescini$^a$ $^d$\\ Daniela Gaglio$^a$ $^c$ $^e$ \quad Marco Vanoni$^a$ $^c$ \quad Lilia Alberghina$^a$ $^c$\quad Giancarlo Mauri$^a$ $^b$\\
\institute{$^a$SYSBIO - Centre for Systems Biology, Piazza della Scienza 2, 20126 Milano, Italy}
\institute{$^b$Universit\`{a} degli Studi di Milano-Bicocca, Dipartimento di Informatica, Sistemistica e Comunicazione\\ Viale Sarca 336, 20126 Milano, Italy}
\email{chiara.damiani@unimib.it, riccardo.colombo@disco.unimib.it, giancarlo.mauri@unimib.it}
\institute{$^c$Universit\`{a} degli Studi di Milano-Bicocca, Dipartimento di Biotecnologie e Bioscienze\\ Piazza della Scienza 2-4, 20126 Milano, Italy}
\email{s.molinari5@campus.unimib.it, daniela.gaglio@ibfm.cnr.it,}
\email{marco.vanoni@unimib.it, lilia.alberghina@unimib.it}
\institute{$^d$Universit\`{a} degli Studi di Milano-Bicocca, Dipartimento di Statistica e Metodi Quantitativi\\ Via Bicocca degli Arcimboldi 8, 20126 Milano, Italy}
 \email{dario.pescini@unimib.it} 
 \institute{$^e$IBFM-CNR,Via Fratelli Cervi 93,Segrate, Milano, Italy}
 \email{daniela.gaglio@ibfm.cnr.it} 
}

\def\titlerunning{Flux Balance Analysis: an ensemble approach}
\def\authorrunning{C. Damiani et al.}
\begin{document}
\maketitle

The search for the molecular basis of oncogenic transformation has long challenged the scientific community. It has been recently recognized~\cite{Hanahan2011} that distinct types of cancer share a few essential functional alterations of normal cell physiology, which taken together define the phenotype of neoplastic diseases.  

The most obvious phenotype of cancer cells is their uncontrolled proliferation, which results from a strong reduction of their response to pro-apoptotic cues and to an enhanced cellular growth. The enhanced cellular growth is, in turn, a consequence of an insensitivity to anti-proliferative signals and it is supported by an extensive rewiring of the cell metabolism to fulfill the increased biosynthetic requirements of the transformed cell~\cite{who2008}. A major metabolic feature of cancer cells is the so-called Warburg effect~\cite{Warburg}, characterized by decreased respiratory activity and enhanced glycolysis-driven lactate production. More recently, it has also been shown that glutamine utilization is significantly stimulated in cancer cells~\cite{Weinberg}.

Thanks to the wide availability of -omics data and well established computational methods, the \textit{cancer metabolic rewiring} (briefly CMR) can be conveniently investigated with a systemic approach combining \textit{in vivo} and \textit{in silico} analyses \cite{Alberghina2012}.\\
A major modeling approach that has been recently applied to cancer is genome-scale metabolic modeling~\cite{Oberhardt}. Being located downstream of the genome, transcriptome and proteome, the metabolome may indeed offer a relatively compact readout of the physiological state of a cell and its perturbation in diseased states~\cite{Zamboni}. However data related to transcriptome or proteome may also be employed to set some constraints of the metabolic model. 

\begin{figure}[ht]
\includegraphics[width=11cm]{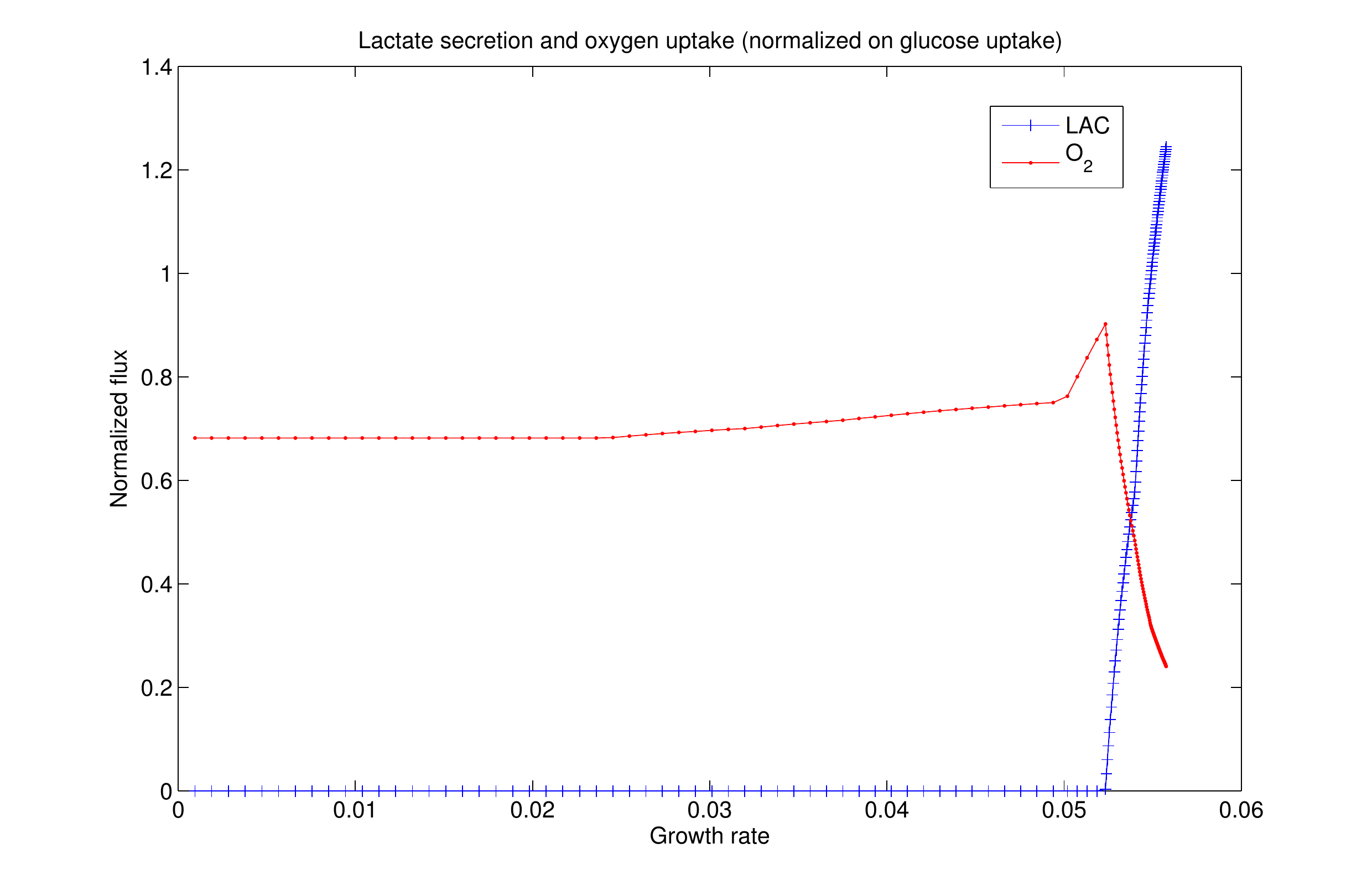}
\centering
\caption{Reproduction of results in \cite{shlomi}: predicted lactate secretion flux (blue lines) and oxygen consumption flux (red lines) for a range of growth rates. The flux value is normalized by the glucose uptake flux of the corresponding growth rate value, which is the flux of the biomass production reaction.}
\vspace{-.5cm}
\label{Figure1}
\end{figure}

In order to perform \textit{in silico} analyses on the CMR, this must be described from a quantitive point of view. In this regard, Shlomi et al.~\cite{shlomi} proposed a mathematical definition of this effect based on the inversion between lactate production (from low to high) and oxygen consumption rates (from high to low), as the growth rate increases, as shown in Figure \ref{Figure1}.

Along with the quantitive definition of the CMR, the description of a model representing the metabolic network under evaluation is crucial as well.
For this purpose, genome-scale and core models provide two alternative modeling approaches widely exploited in literature. On one side, large networks apparently require less assumptions concerning the reactions that need to be included in the model but, despite of this, make the interpretation and the understanding of the simulation outcomes not always straightforward. On the other side, small scale models need much more assumptions to properly define the set of reactions, but they are able to highlight the emergent properties of the system under evaluation in a simpler way.

\textit{In silico} analyses of the metabolic rewiring leading to enhanced growth are mainly based on stoichiometric considerations and on enzymatic costs \cite{heiden,enzymeCosts}.
In particular, Flux Balance Analysis (FBA) \cite{Orth2010} is a constraint based method that calculates the flow of metabolites (i.e. the flux) through a metabolic network by relying on: i) stoichiometric constraints ii) the assumption that metabolic networks will reach a quasi-steady state iii) an objective function (OF). The steady state assumption reduces the system to a set of linear equations which is then solved to find a flux distribution that satisfies the steady state condition subject to the stoichiometry constraints while maximizing the flux of one of more specific reactions (the objective function). The solution space identified by the OF may be further restricted by specifying the boundaries of the flux through any particular reaction, by introducing constrains derived e.g. from âomicsâ analyses \cite{kauffman}.
Because of its requirements, which overcome some drawbacks of mechanistic modeling (e.g. need for parameters) and network-based analyses (e.g. lack of stoichiometric knowledge), FBA is particularly suitable to be applied to either genome wide reconstructions \cite{shlomi, Thiele2013} or small scale core models \cite{Resendis2010, Si2009}.

As long as abnormal proliferation is concerned, the function to be optimized takes typically the form of a pseudo-reaction representing the conversion of biomass precursors into biomass.
However, we observed that the results of FBA studies are generally largely sensitive to variations in the definition of such pseudo-reaction, as well as to small variations in the flux constraints, implying that the appearance of a metabolic switch (i.e. the preference of cancer cells for glycolysis with respect to respiration) may result from ad hoc assumptions.

For the aforementioned reasons, we addressed the issue of enhanced growth phenotype of cancer cells by comparing results emerging from FBA analysis of metabolic networks at different levels of abstraction. Besides, as the choice of the OF highly affects the computational outcomes of the FBA method and because every author gives a different formulation of the OF, based on the target to maximize and on the organism under investigation, during this phase, simulations have been conducted with different OFs inspired by literature \cite{shlomi,Selvarasu2010,Resendis2010}.

As a first step, we developed a simplified core model (40 metabolites and 59 reactions) of the main metabolic pathways involved in neoplastic transformation (glycolysis, TCA cycle, OXPHOS, pentose phosphate and glutaminolysis)
using as reference the experimental results described in \cite{Gaglio2011}.
Our core model has been manually built including exchange reactions to maintain the quasi-steady state assumption of FBA and has been tested in order to verify its feasibility i.e. its ability to find a flux distribution different from the trivial one (all fluxes equal to zero).\\
Nevertheless, the outcome of any tested OF has not been able to reproduce the CMR.

In order to cope with this this inability, we propose here a complementary approach inspired by the ensemble approach to the study of biological networks suggested by Kauffman \cite{ensemble}.
The idea is to find one or more ensembles of models defined by specific structural constraints, whose generic properties statistically match those of real cells and organisms. \\
The suggested method consists in studying the general behavior of ensembles of randomly constructed biological networks hypothesizing their ability to display properties that are biologically plausible. The underlying assumption is that real biological networks are somehow typical members of a class, or ensemble, of networks which selection has modified to some degree.
%
%
%
%
%
%
\begin{figure}[ht]
\centering
\begin{tabular}{c}
\begin{tabular}{c c}
\includegraphics[width=6cm]{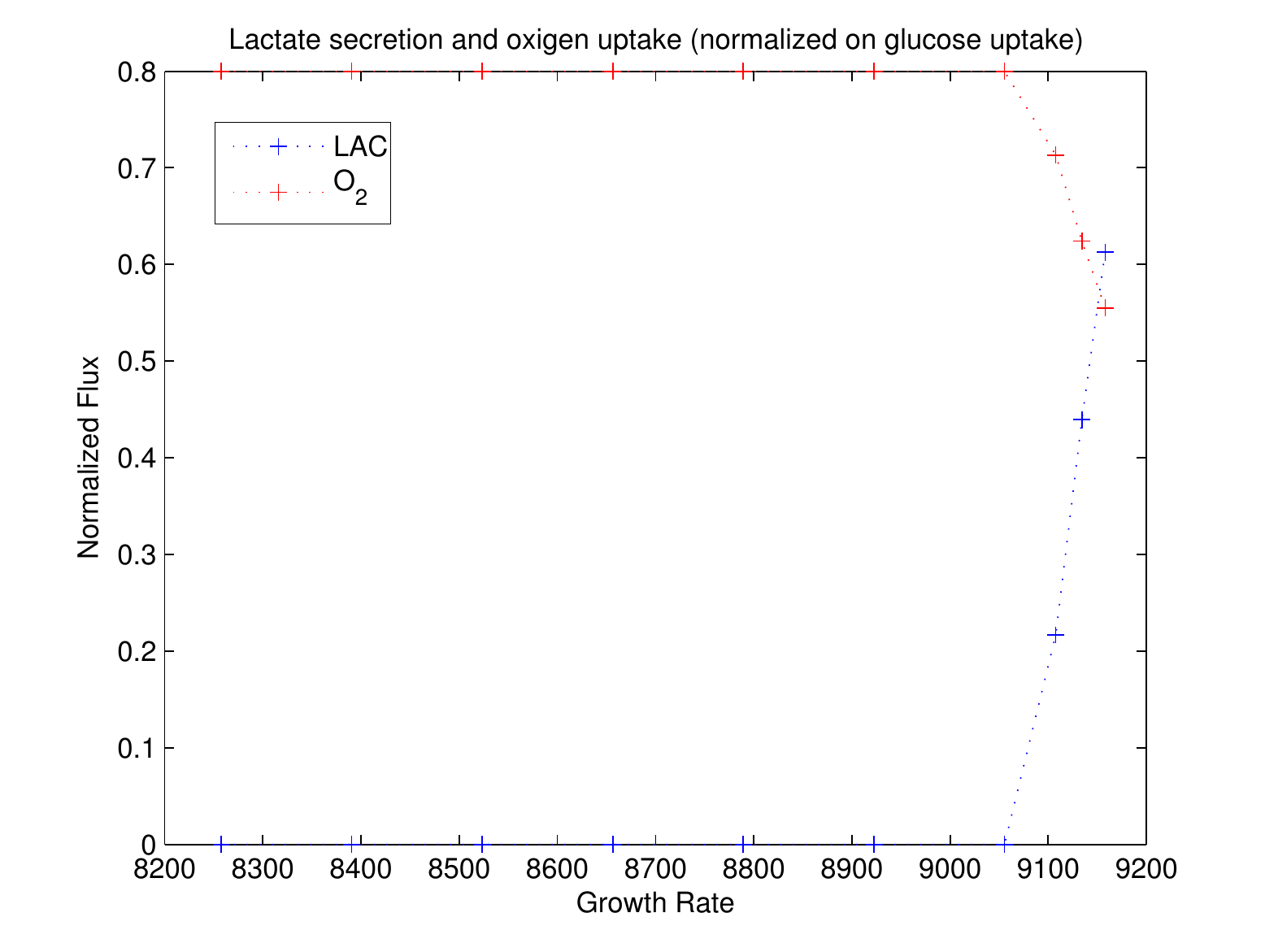} & \hspace{-.5cm} \includegraphics[width=6cm]{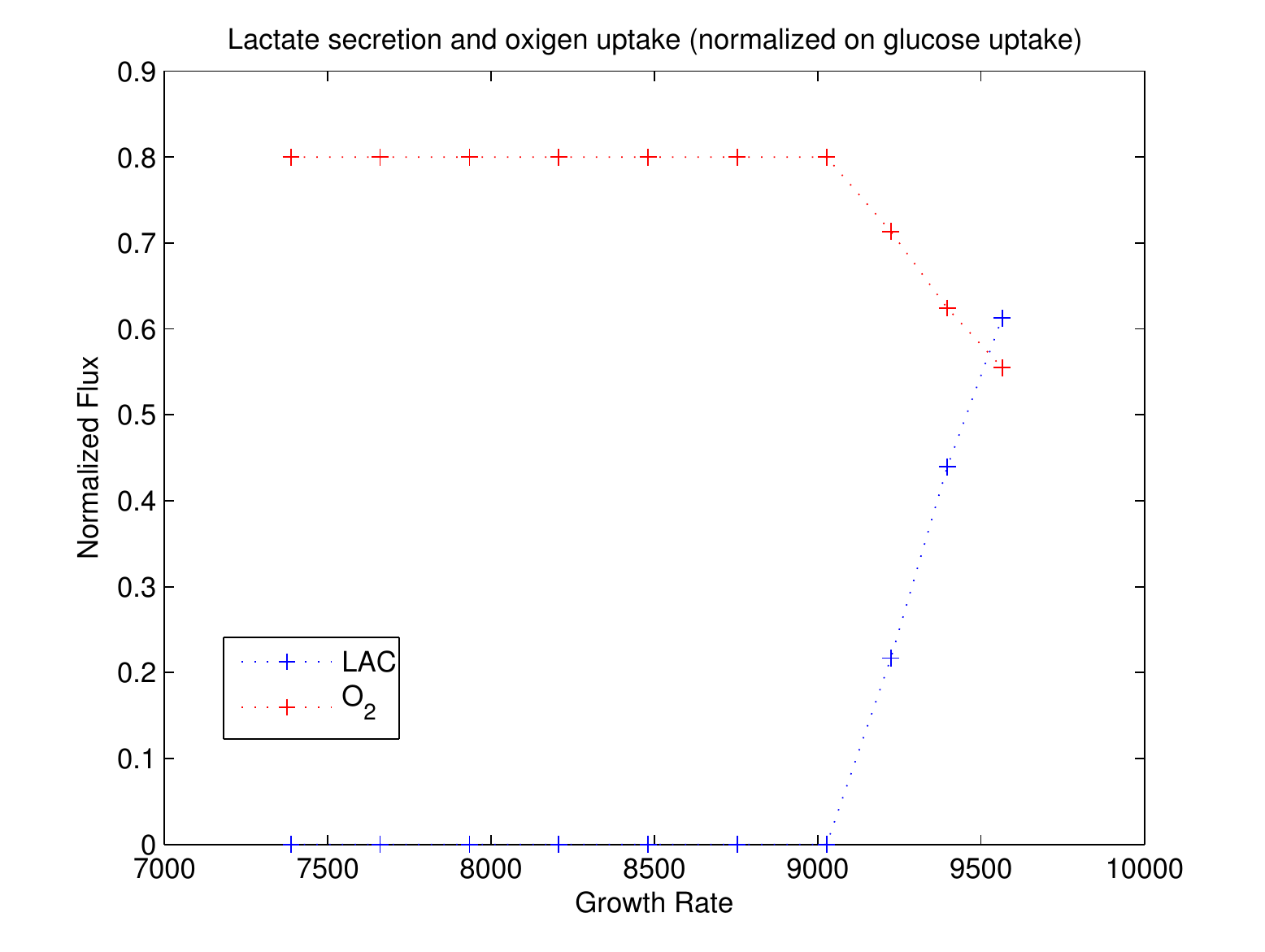}\\ 
\includegraphics[width=6cm]{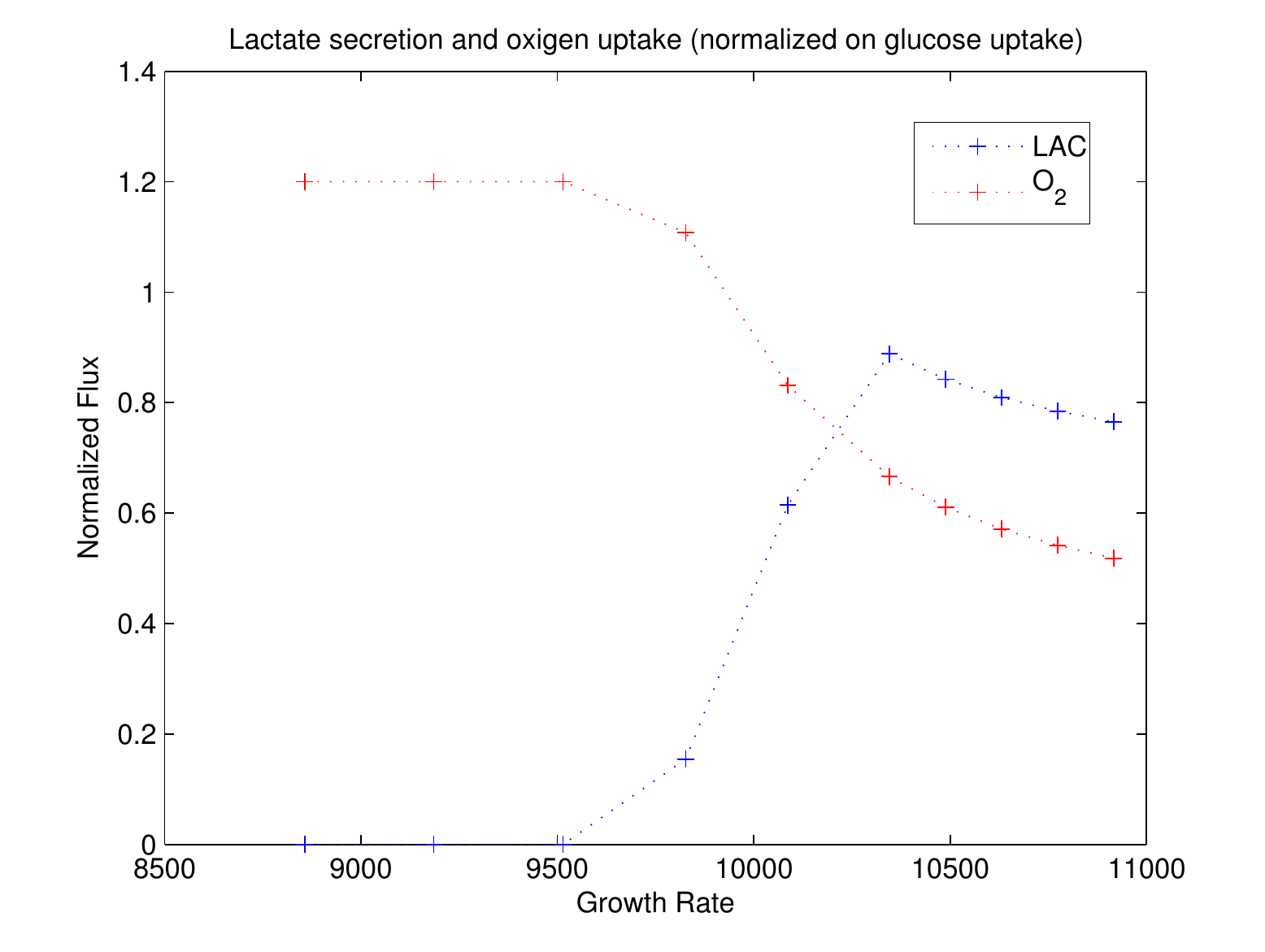} & \hspace{-.5cm} \includegraphics[width=6cm]{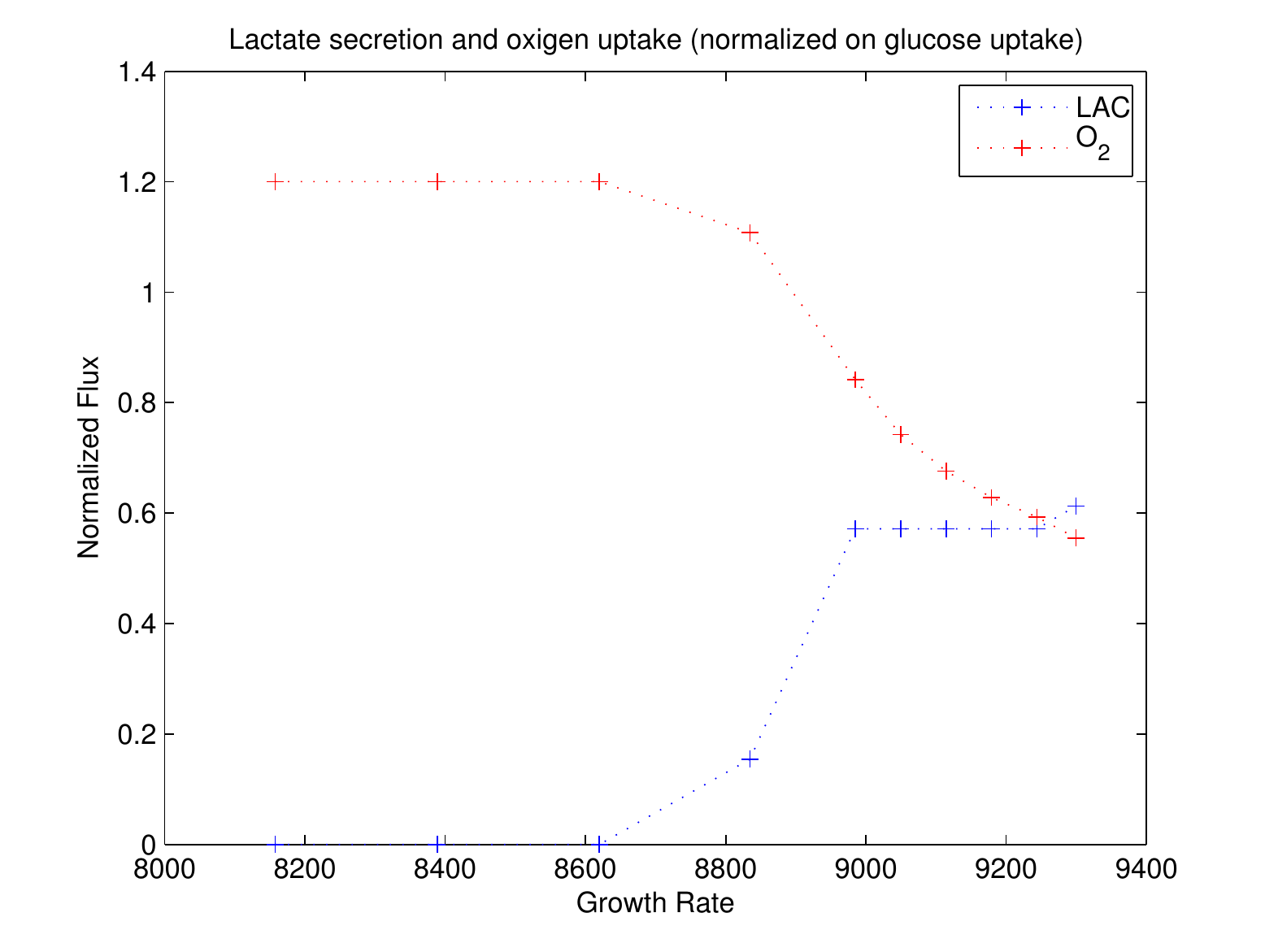}\\ 
\vspace{-1.1cm}
\end{tabular}\\
\hspace{-1cm}
\vspace{-1cm}
\includegraphics[width=17cm]{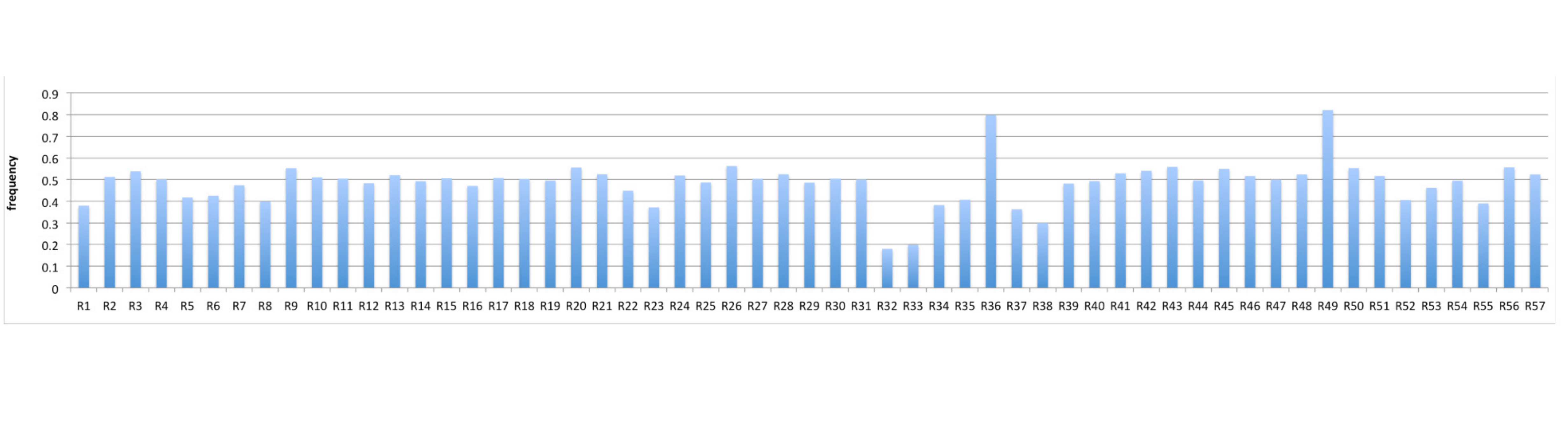}\\ 
\end{tabular}
\caption{{\bf Top:} Examples of CMR obtained form our core model with randomly generated succesfull OFs. {\bf Bottom:} Frequency at which each reaction appears in the set of successful functions. Results are obtained from 100000 randomly generated OFs which gave rise to 84 succesfull runs.}
\label{Figure2}
\end{figure}
Therefore, the generic properties of the ensemble members may be able to provide insight into the structure, logic and dynamics of real networks; or, at least, they can serve as useful null hypotheses about what we would expect to find and direct further experimental work.\\
A first attempt in this direction has been made by exploring the space of the possible objective functions (stoichiometric matrix and  constraints being invariant), by randomly generating them, and by selecting those leading to the expected result.\\
Specifically, a particular instance of random OF is obtained as follows: a subset of reactions within the model is drawn, then each reaction in the subset is assigned a random objective coefficient, which weights the flux of such reaction with respect  the OF value. \\
In order to evaluate the so obtained OF, the expected result must be quantitatively defined. To this end, we focused on the narrow description of CMR proposed by Shlomi et al. \cite{shlomi} (see Figure \ref{Figure1}, left side): different growth rates values are obtained by optimizing the model at different levels of glucose uptake and the relative flux distributions are then investigated.
If, in correspondence of the lowest (highest) growth rate, the oxygen uptake flux overcomes (is beneath) that of lactate secretion, then we consider that a metabolic switch has occurred.

The methodology has been tested on the core metabolic network previously described, revealing that a non-negligible fraction of randomly assigned OFs leads to the CMR, accordingly to the definition given above (some examples are shown in Figure \ref{Figure2}, top).\\
An ensemble of successful functions has therefore been obtained and must then be analyzed in order to detect some generic properties that an OF must possess for the metabolic rewiring to be observed.
For instance, we may want to know which are the reactions that are supposed to participate to the OF. In this regard, Figure \ref{Figure2} (bottom) represents the frequency at which each reaction appears in the set of successful functions. One can notice that a few reactions outperform all the others, suggesting their pivotal role in the onset of the CMR.

This result will be further validated by applying the same procedure to the genome-wide models Recon 1 \cite{Duarte2007} and Recon 2 \cite{Thiele2013} and to an expansion of our core model. \\
Furthermore, in the near future, deriving inspiration from the domain of metabolic engineering \cite{Rocha2007}, we will define and implement an evolutionary algorithm in order to find the objective function (in terms of components and coefficients) that best fits data from ``wet'' experiments.\\
Moreover, investigations will include an accurate comparison and discussion of results obtained from a partial modification of the genome scale model based on \cite{shlomi} and from our developed core model.
Finally, the integration of constraint and mechanism-based models will be developed in order to understand the enhanced proliferation of cancer cells under different perspectives.


\bibliographystyle{eptcsini}
\bibliography{WivaceFBA}
\end{document}